# Acoustooptic operation of optical vortex beams

**Iryna Martynyuk-Lototska, Yurii Vasylkiv, Taras Dudok, Ihor Skab, and Rostyslav Vlokh**

*Vlokh Institute of Physical Optics, 23 Dragomanov Street, 79005, Lviv, Ukraine*

**Using acoustooptic (AO) cells based on TeO$_2$ crystal and silica glass, we have experimentally shown for the first time that the intensity profile and the phase structure of the vortex beam are preserved under AO Bragg diffraction. As a result, the vortex beam can be deflected due to AO diffraction, whereas the acoustooptically operated vortex beams can be efficiently used in such novel branches of optical technology as optical trapping and controlled addressing of the beams with different orbital angular momentums**.

Optical vortices bearing nonzero orbital angular momentum (OAM) can be used in different branches of optical technologies, e.g. in quantum computing [1], quantum communications, beam focusing below diffraction limit [2], and microparticle manipulation [3]. Simultaneous availability of OAM ($l=0,\pm 1,\pm 2...$) and spin angular momentum ($\sigma^{+,-}$) in an optical beam increases the number of possible states $|\sigma^{+,-},l\rangle$ in which the information can be encoded [4]. Then any photon can carry arbitrarily large amount of information distributed over its spin and orbital quantum states [5]. This is why controlled addressing of the beams with different quantum states has become an important problem, when using photons as carriers of encoded information. In addition, spatial operation of the vortex beams acquires a fundamental character if one deals with optical trapping of microparticles. Mechanical methods such as gimbal-mounted mirrors [6], computer-controlled galvanomirrors [7] and piezoelectric mirrors [8], along with electrooptic and acoustooptic (AO) methods, are among the main techniques used for spatial operation of optical beams with the purpose of microparticle trapping [9–11]. Note that all of the methods mentioned above deal with the Gaussian beams. It is also known that some of the methods for optical trapping are based on the effect of radiation pressure. They are associated with gradient forces [12, 13] (including the pressure of evanescent field [14] and that appearing in the vortex beam [3]) or, alternatively, photophoretic forces [15, 16]. Each of the methods related to different types of optical beams reveals both advantages and drawbacks. For example, the trapping based on the radiation pressure and the Gaussian beams is applicable when manipulating with non-absorbing dielectric particles characterized by relatively high refractive indices, whereas the photophoretic forces are limited to trapping of absorbing particles only. The beams that bear nonzero OAM can be used for nondestructive manipulation of absorbing particles and the particles with low refractive indices, which is important for many biologic applications [17].

The most common technique employed to deflect the optical beams mentioned above is based upon AO effect. Here the beams can easily be deflected via changes in the acoustic wave frequency, while the efficiency of Bragg diffraction can be controlled by the acoustic power (see, e.g., Ref. [18]). Moreover, two consecutive AO cells make it possible to implement 2D deflection and addressing of microparticles to any desirable places, with high enough spatial resolutions.

As a matter of fact, the effect of exchange by the angular momentums between the acoustic and optical beams under AO diffraction has already been successfully demonstrated (see Refs. [19, 20]). However, the AO diffraction itself represents a complicated process which can be accompanied by changing degree of coherence of Bragg-diffracted optical waves [21]. In principle, the latter can lead to instability or even destruction of the phase structure of a helical mode. Hence, it would be vital to clarify whether the vortex beam can be deflected by AO gratings without destruction of the phase structure of the vortex. This is of primary importance for the quantum communications, quantum computing and microparticle manipulation. In this Letter, we demonstrate that the vortex beams can indeed be deflected by AO cells such that the phase distribution in their cross sections is preserved.

In our experiments we have used two different AO cells, an AO deflector based on TeO$_2$ crystal and a Q-switch made of fused silica. We have chosen TeO$_2$ because it is one of the most efficient AO materials often used for light deflecting, with the AO figure of merit being as high as $1200\times 10^{-15}$ s$^3$/kg

[22]. This material is also convenient for AO operation of microparticle trapping. Owing to its high acoustic-wave velocity, silica glass is characterized by the AO rise times ~ 0.5 μs for the light beam diameter 3 mm, which is important for controlled addressing of OAM-bearing beams.

Our paratellurite sample was fabricated in the shape of parallelepiped, with the faces perpendicular to the directions [001], [110] and [1$\bar{1}$0], and the corresponding thicknesses equal to 12, 10 and 9 mm, respectively. The optical vortex beam propagated close to the [001] axis, while the slow shear acoustic wave with the polarization parallel to [1$\bar{1}$0] and the velocity equal to 616 m/s propagated along the direction [110]. The acoustic bandwidth of our deflector amounted to 60–30 MHz. The acoustic wave was excited with a standard LiNbO$_3$ piezoelectric transducer. The working AO element of our Q-switch made of silica glass had the length 44 mm (measured along the light propagation direction), the width 34–38 mm (along the direction of acoustic wave vector) and the height 11 mm. The longitudinal acoustic wave propagating with the velocity 5960 m/s was excited using a thick piezoelectric plate made of LiNbO$_3$, with the central frequency 50 MHz. The Bragg angle was equal to ~ 7.0 and 0.3 deg for the cases of paratellurite and quartz cells, respectively.

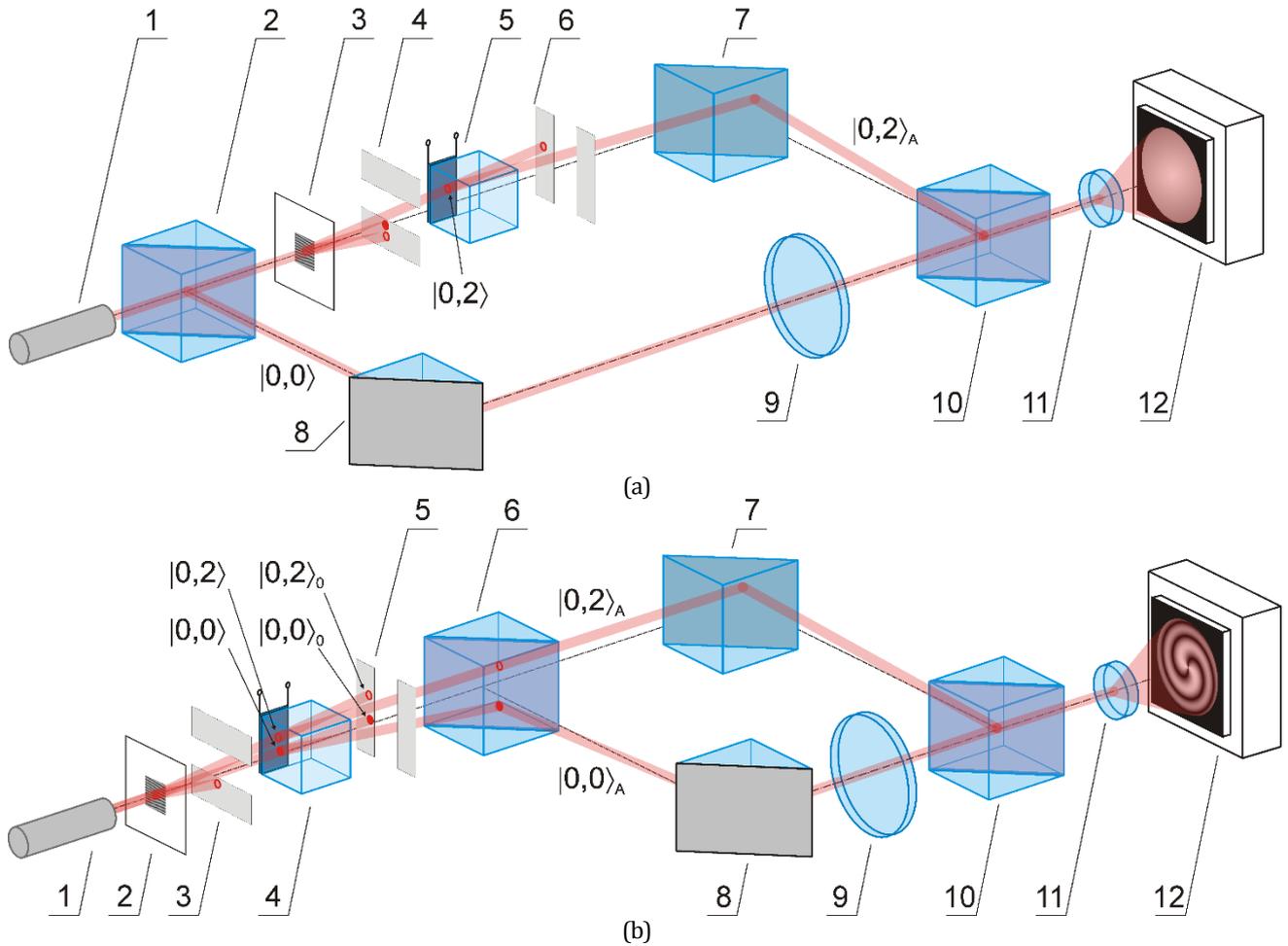

Fig. 1. Experimental set-ups used for studying interference of an acoustooptically diffracted vortex beam with a Gaussian spherical beam: (a) (1) a He–Ne laser (the wavelength 632.8 nm), (3) a computer-synthesized hologram, (4) and (6) diaphragms, (5) an AO cell, (2) and (10) beam splitters, (7) and (8) reflection prisms, (9) an optical lens with the focal length 50 cm, (11) an objective lens, and (12) a CCD camera;
(b) (2) a computer-synthesized hologram, (3) and (5) diaphragms, (4) an AO cell, (6) and (10) beam splitters, (1, 7-12) the same as in panel (a).

In panels (a) and (b), the computer-synthesized hologram and the AO cell are inserted into the sample arm of a Mach–Zehnder interferometer and in front of this interferometer, respectively.

Our experimental set-up provided two alternative configurations that differed by where a computer-synthesized hologram and an AO cell were put in the optical scheme. In the first case they were placed into the sample arm of a Mach–Zehnder interferometer (see Fig. 1a), whereas the second case differed in that the said components were located in front of the interferometer (see Fig. 1b). A linearly polarized light was emitted by a He–Ne laser with the coherence length of about 0.3 m. The optical vortex beam was generated with the aid of a computer-synthesized hologram. According to the technique described in Ref. [23], the latter was fabricated with the resolution 3048 dpi, using a standard transparent film. The hologram included a doubly charged topological defect of fringe ordering. Then the diffracted vortex beams had even charges. To be specific, we used the beams with the vortex charge equal to ±2, which revealed the highest intensity among the diffracted beams.

The vortex beam denoted conventionally as $|0,2\rangle$ falls upon the AO cell under the Bragg angle and the first-order diffracted beam interferes with the reference spherical Gaussian beam referred to as $|0,0\rangle$. For the case of configuration depicted in Fig. 1a, we have observed no interference with either TeO$_2$ or silica AO cells. Nonetheless, a characteristic doughnut shape of the intensity profile of the beam is still preserved behind the AO cell (see Fig. 2a). Any natural experimental manipulations that spring to mind, like changing the acoustic wave frequency and the acoustic power, or replacing the AO cells, have not resulted in the appearance of interference pattern. At the same time, Fig. 2b testifies that the zero-order diffraction AO maximum does interfere with the spherical Gaussian beam, thus generating double spiral interference fringes peculiar for the interaction of spherical Gaussian beams with the beams bearing doubly charged optical vortices.

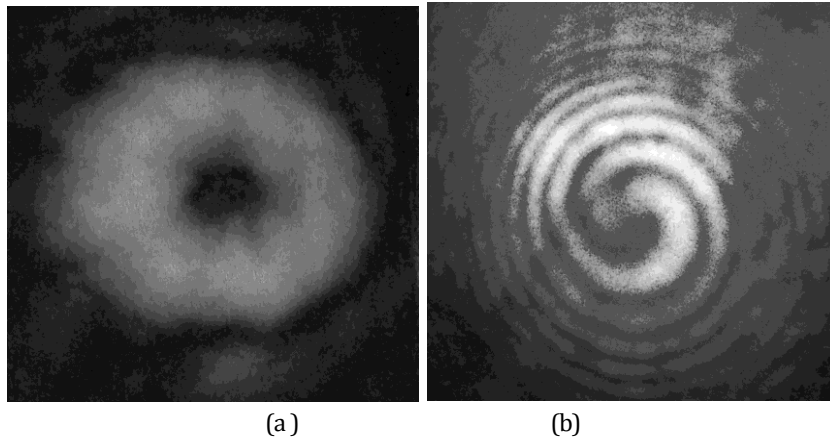

(a) (b)

Fig. 2. Doughnut intensity shapes observed for the first-order acoustooptically diffracted beam (a) and interference pattern appearing due to interference of the zero-order diffracted vortex beam with the spherical reference beam (b). Here we deal with the Bragg diffraction in TeO$_2$ cell at the acoustic frequency 100 MHz.

The above experimental facts hint that the first-order diffracted vortex beam is not coherent with the reference one, although the zero-order and reference beams remain coherent. The main difference between the zero- and first-order diffracted beams consists in their Doppler shifts, i.e. the changes in the diffracted optical wave frequency equal to the acoustic frequency. Despite of the fact that the Doppler shift is very small (about $2 \times 10^{-5}$ % in our case), it must be enough to violate the exact interference conditions.

The situation can be improved using the experimental geometry of the second kind, as shown in Fig. 1b. In the latter experiment, both the zero-order diffracted Gaussian beam (denoted now as a beam $|0,0\rangle$) and the first-order diffracted beam (referred to as a vortex beam $|0,2\rangle$), which is produced by diffraction at the computer-synthesized hologram, diffract acoustooptically at the AO cell. Four beams appear due to the diffraction, two zero-order beams (a Gaussian beam $|0,0\rangle_0$ and a vortex beam $|0,2\rangle_0$) and two first-order diffracted ones (an acoustooptically diffracted spherical Gaussian beam $|0,0\rangle_A$ and a vortex beam $|0,2\rangle_A$). Here the Doppler shifts for the beams $|0,0\rangle_A$ and $|0,2\rangle_A$ are the same. After this, both of the beams $|0,0\rangle_A$ and $|0,2\rangle_A$ are directed into the interferometer in such a way that the first one plays the role of a reference beam and the second one serves as a probe beam. As seen from Fig. 3, these optical beams interfere successfully, since they have the same Doppler shifts. These experimental results have been successfully reproduced for the both of our AO cells. This proves that the vortex beams preserve their coherence and the structure of their phase profiles under AO diffraction.

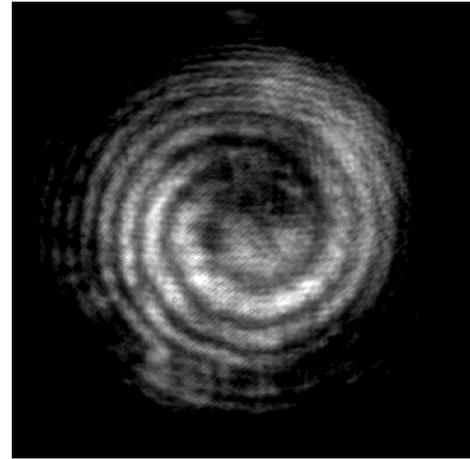

Fig. 3. Spiral interference patterns appearing due to interference of acoustooptically diffracted spherical Gaussian beam and the vortex beam (Bragg diffraction using our AO TeO$_2$ cell and the acoustic frequency 100 MHz).

Our next task is to give evidence that changing the frequency of the acoustic wave can be efficiently used for spatial operation of the vortex beams. For this aim, two acoustic waves with different frequencies, 85 and 95 MHz, were excited in our TeO$_2$-based AO deflector. Then the lower frequency was raised gradually from 85 up to 95 MHz. As seen from Fig. 4a–d, the two vortices appearing under these conditions are separated by still less angular distances (0.85, 0.34, 0.25 and 0 deg, respectively). This result suggests that AO deflection of a single vortex beam, or even several vortex beams, enables one to operate efficiently the process of microparticle manipulation and facilitates controlled addressing of the beams that bear different OAMs.

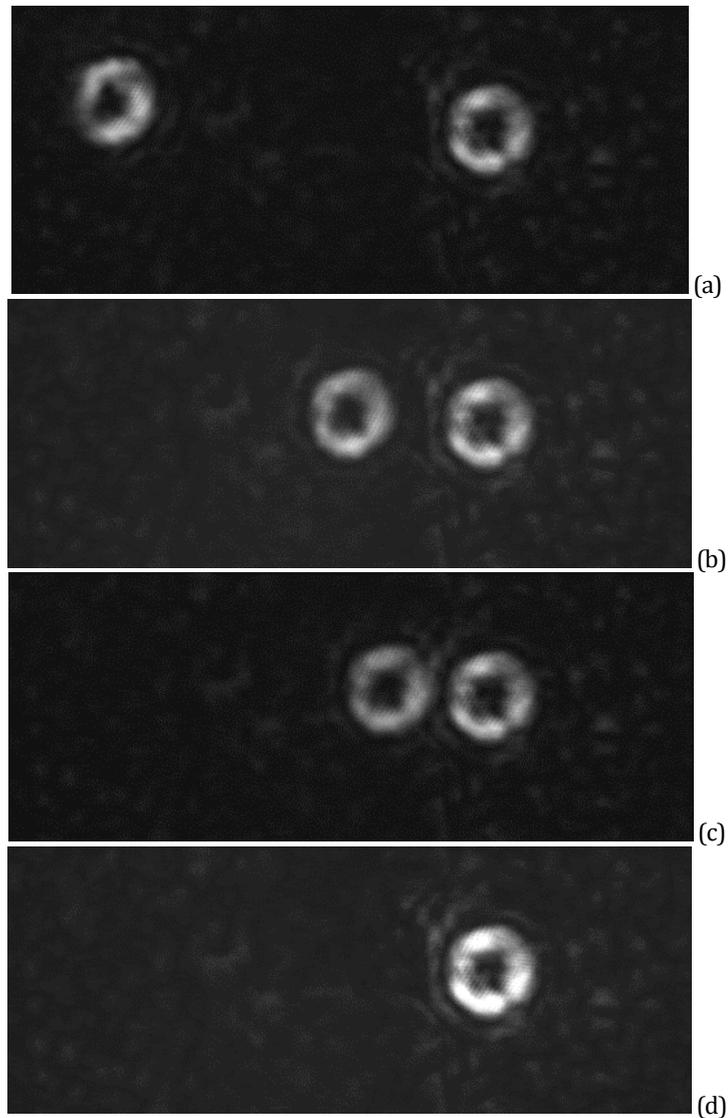

Fig. 4. Bragg diffraction of doubly charged optical vortex at the acoustic waves with the frequencies 85 and 95 MHz (a), 91 and 95 MHz (b), 92 and 95 MHz (c), and 95 and 95 MHz (d) ( Bragg diffraction using our AO TeO$_2$ cell).

Summing up, we have demonstrated for the first time that the phase structure of the vortex beams is preserved under the AO Bragg diffraction. As a result, acoustooptically operated vortex beams can be efficiently used in such novel optical technologies as the optical trapping or the controlled addressing of beams with different OAMs.

**References**

1. M. A. Nielsen and I. L. Chuang, *Quantum computation and quantum information* (Cambridge University Press, 2000).
2. T. Brunet, Jean-Louis Thomas, and R. Marchiano, Phys. Rev. Lett. **105**, 034301 (2010).
3. K. T. Gahaganand and G. A. Swartzlander, Opt. Lett. **21**, 827 (1996).
4. L. Allen, M. W. Beijersbergen, R. J. C. Spreeuw, and J. P. Woerdman, Phys. Rev. A. **45**, 8186 (1992).
5. G. Molina-Terriza, J. P. Torres, and L. Torner, Phys. Rev. Lett. **88**, 013601 (2001).
6. E. Fallman and O. Axner. Appl. Opt. **36**, 2107 (1997).
7. K. Sasaki, M. Koshioka, H. Misawa, N. Kitamura, and H. Masuhara, Jpn. J. Appl. Phys. **30**, L907 (1991).
8. C. Mio, T. Gong, A. Terray, and D. W. M. Marr. Rev. Sci. Instrum. **71**, 2196 (2000).
9. K. Visscher, S. P. Gross, and S. M. Block, IEEE J. Select. Topics Quant. Electron. **2**, 1066 (1996).
10. A. L. Oldenburg, S.-J. Moon, K. M. Kasi, T. Kim, C. Ho, R. Timp, H. Choi, V. I. Gelfand, J. Roland, K. Kim, S. A. Boppart, and G. L. Timp. Proc. SPIE. **4962**, 249 (2003).
11. I. A. Martínez and D. Petrov, **51**, Appl. Opt. 5522 (2012).
12. A. Ashkin, Phys. Rev. Lett. **24**, 156 (1970).
13. A. Ashkin, J. M. Dziedzic, J. E. Bjorkholm, and S. Chu, Opt. Lett. **11**, 288 (1986).
14. S. Kawata and T. Sugiura, Opt. Lett. **17**, 772 (1992).
15. Y. I. Yalamov, V. B. Kutukov, and E. R. Shchukin. J. Colloid Interface Sci. **57**, 564 (1976).
16. B. Redding, S. C. Hill, D. Alexson, C. Wang, and Y.-L. Pan, Opt. Express. **23**, 3630 (2015).
17. K. Svoboda and S. M. Block, Ann. Rev. Biophys. Biomol. Struct. **23**, 247 (1994).
18. N. Savage, Nature Photonics. **4**, 728 (2010).
19. P. Z. Dashti, F. Alhassen, and H. P. Lee, Phys. Rev. Lett. **96**, 043604 (2006).
20. V. N. Belyi, P. A. Khilo, N. S. Kazak, and N. A. Khilo, J. Opt. **18**, 074002 (2016)
21. C.-W. Tarn, J. Opt. Soc. Am. A. **16**, 1395 (1999).
22. http://www.aaoptoelectronic.com/1.aspx
23. A. V. Carpentier, H. Michinel, J. R. Salgueiro, and D. Olivieri, Am. J. Phys. **76**, 916 (2008).